\documentclass[aip,jcp,preprint,dvips]{revtex4-1}
\usepackage{amsmath}
\usepackage{amssymb}
\usepackage{graphicx}
\usepackage[usenames,dvipsnames]{color}

\begin{document}
\title{Quantum Heat Current under Non-perturbative and Non-Markovian Conditions: Applications to Heat Machines}

\author{Akihito Kato}
\email{kato@kuchem.kyoto-u.ac.jp}
\author{Yoshitaka Tanimura}
\email{tanimura@kuchem.kyoto-u.ac.jp}
\affiliation{Department of Chemistry, Graduate School of Science, Kyoto University, Kyoto 606-8502, Japan}
\date{\today}

\begin{abstract}
We consider a quantum system strongly coupled to multiple heat baths at different temperatures.
Quantum heat transport phenomena in this system are investigated using two definitions of the heat current,
one in terms of the system energy, and the other in terms of the bath energy.
When we consider correlations among system-bath interactions (CASBI)
-- which have a purely quantum mechanical origin --
the definition in terms of the bath energy becomes different.
We found that CASBI are necessary
to maintain the consistency of the heat current with thermodynamic laws
in the case of strong system-bath coupling.
However, within the context of the quantum master equation approach,
both of these definitions are identical.
Through a numerical investigation,
we demonstrate this point for a non-equilibrium spin-boson model
and a three-level heat engine model
using the reduced hierarchal equations of motion approach
under strongly coupled and non-Markovian conditions.
We observe cyclic behavior of the heat currents and the work performed by the heat engine, and we find that their phases depend on the system-bath coupling strength. Through consideration of the bath heat current, we show that the efficiency of the heat engine decreases as the strength of the system-bath coupling increases, due to the CASBI contribution. In the case of a large system-bath coupling, the efficiency decreases further if the bath temperature is {increased}, even if the ratio of the bath temperatures is fixed,
due to the discretized nature of energy eigenstates.
This is also considered to be a unique feature of quantum heat engines.
\end{abstract}
\maketitle

\section{Introduction}\label{sec:intro}
Recent progress in the control and measurement of small-scale systems
provides the possibility of examining the extension of thermodynamics \cite{Ritort,Seifert,Campisi,Brandao}
and the foundation of statistical mechanics \cite{Trotzky,Gemmer} in nano materials.
In particular, elucidating how such purely quantum mechanical phenomena
as quantum entanglement and coherence are manifested in thermodynamics
has become a topic of growing interest for the past two decades. \cite{Nori0, Nori, Nori1,Brunner,Chotorlishvili,Uzdin,Lostaglio,Goold}
Investigations of these types of phenomena may provide new insights into our world
and aid in the development of quantum heat machines operating beyond classical bounds.
Such developments could lead to more efficient methods of energy usage.

Quantum heat transport problems involving quantum heat machines
have been studied with approaches developed through application of open quantum dynamics theory.
Quantum master equation (QME) approaches are frequently used to investigate the dynamics of quantum heat machines. \cite{Kosloff14,Gelbwaser15}
Although the QME approach is consistent with the laws of thermodynamics, \cite{Alicki,Kosloff13,Breuer}
its applicability is limited to cases in which
the system-bath interaction is treated as a second-order perturbation
and the Markov approximation is employed.
Thus, these investigations have been carried out only in weak coupling regimes.

Recent theoretical and experimental works \cite{Huelga,Engel} have demonstrated the importance of the interplay between the quantum nature of systems and environmental noise.
For example, it has been shown that the optimal conditions for excitation energy transfer in light-harvesting complexes
is realized in the non-perturbative, non-Markovian regime, in which a description beyond perturbative approaches
is essential to properly understand the quantum dynamics displayed by the system.
To this time, the approaches used to study this regime include
the QME employing a renormalized system-plus-bath Hamiltonian derived with the polaron transformation \cite{Gelbwaser}
or the reaction-coordinate mapping, \cite{Strasberg,Newman} the functional integral approach, \cite{Carrega}
the non-equilibrium Green's function method, \cite{Esposito15PRL,Esposito15PRB,Nitzan} and the stochastic Liouville-von Neumann equation approach.\cite{Schmidt}
In most cases, however, such attempts are limited to the nearly Markovian case, slow driving cases, or the investigation of the short-time behavior.
In the present study, we employ the hierarchal equations of motion (HEOM) approach \cite{Tanimura88,Ishizaki05,Tanimura06,Tanimura14, Tanimura15, Kato15}
to investigate heat transport and quantum heat engine problems.
This approach allows us to treat systems subject to external driving fields in a numerically rigorous manner under non-Markovian and
{non-perturbative system-bath coupling conditions. We must chose the definition of heat current carefully, however, to satisfy various thermodynamic requirements,
for example, to have right thermal equilibrium limit. \cite{Esposito15PRB}
While several researchers have studied a role of heat current between subsystems, \cite{Tannor,Segal,Castro}
which are introduced by partitioning a many-body system such as chain models,
here we consider the heat current between the system and baths.
Thus} we employ two definitions of the heat current, one in terms of the system energy (system heat current), and the other in terms of the bath energy (bath heat current). Both of these definitions are frequently used in the literature {for varieties of systems involving chain models, here} we carefully examine the validity of these definitions {for a non-equilibrium spin-boson model and a three-level heat engine model} in the case of the strong system-bath coupling,
because the existence of a non-commuting inter-bath coupling through the system
contributes to the heat current even in the steady state case. 
This effect has not been investigated in previous studies.

The organization of this paper is as follows.
In Sec. \ref{sec:formula}, we introduce the two definitions of the heat current that we investigate, the system heat current (SHC), and the bath heat current (BHC).
In Sec. \ref{sec:thermodynamics}, we present the first and the second laws of thermodynamics as obtained through consideration of the BHC.
In Sec. \ref{sec:reduced}, we analytically derive reduced expressions for the SHC and BHC.
In Sec. \ref{sec:heom}, we explain the HEOM approach and demonstrate a method employing it to calculate the SHC and BHC numerically in a rigorous manner.
In Sec. \ref{sec:numerics}, we apply our formulation to a non-equilibrium spin-boson model and a three-level heat engine model.
Through numerical investigation of the HEOM, we investigate the cycle behavior of the quantum heat engine under a periodic driving field.
It is shown that the efficiency of the heat engine decreases as the system-bath coupling increases.
Section \ref{sec:conclusion} is devoted to concluding remarks.

\section{System Heat Current and Bath Heat Current}\label{sec:formula}
We consider a system coupled to multiple heat baths at different temperatures.
With $K$ heat baths, the total Hamiltonian is written
\begin{align}
\hat{H}(t)
= \hat{H}_\mathrm{S}(t)
  + \sum_{k=1}^K \left(
  \hat{H}_\mathrm{I}^k + \hat{H}_\mathrm{B}^k \right),
\label{eq:H_total}
\end{align}
where $\hat{H}_\mathrm{S}(t)$ is the system Hamiltonian,
whose explicit time dependence originates from the coupling with the external driving field.
The Hamiltonian of the $k$th bath and the Hamiltonian representing the interaction between the system and the $k$th bath are given by
$\hat{H}_\mathrm{B}^k = \sum_j \hbar \omega_{k_j} \hat{b}_{k_j}^\dagger \hat{b}_{k_j}$
and $\hat{H}_\mathrm{I}^k = \hat{V}_k \sum_{j} g_{k_j}( \hat{b}_{k_j}^\dagger + \hat{b}_{k_j})$, respectively,
where $\hat{V}_k$ is the system operator that describes the coupling to the $k$th bath.
Here, $\hat{b}_{k_j}, \hat{b}_{k_j}^\dagger, \omega_{k_j}$ and $g_{k_j}$ are
the annihilation operator, creation operator, frequency, and coupling strength for the $j$th mode of the $k$th bath, respectively.
Due to the bosonic nature of the bath, all bath effects on the system are determined by the noise correlation function,
$C_k(t) \equiv \langle \hat{X}_k(t) \hat{X}_k(0) \rangle_\mathrm{B}$,
where $\hat{X}_k \equiv \sum_j g_{k_j}( \hat{b}_{k_j}^\dagger + \hat{b}_{k_j} )$ is the collective bath coordinate of the $k$th bath
and $\langle \ldots \rangle_\mathrm{B}$ represents the average taken with respect to the canonical density operator of the baths.
The noise correlation function is expressed in terms of the bath spectral density, $J_k(\omega)$, as
\begin{align}
C_k(t)
= \int_0^\infty d\omega \, \frac{ J_k(\omega) }{ \pi }
  \left[ \coth\left( \frac{\hbar\omega}{2k_BT_k} \right) \cos(\omega t)
  - i \sin(\omega t) \right],
\label{eq:noise}
\end{align}
where $J_k(\omega) \equiv \pi \sum_j g_{k_j}^2 \delta(\omega - \omega_{k_j})$,
and $T_k$ is the temperature of the $k$th bath.

For the system described above, we derive two rigorous expressions for the heat current,
which are convenient for carrying out simulations of reduced system dynamics.
One of these expressions is derived through consideration of conservation of the the system energy,
and for this reason, we call it the "system heat current" (SHC).
This current is defined as
\begin{align}
\frac{d}{dt}\langle \hat{H}_\mathrm{S}(t) \rangle - \dot{W}(t)
= \sum_{k=1}^K \dot{Q}_\mathrm{S}^k(t),
\end{align}
where $\dot{W}(t) \equiv \langle (\partial \hat{H}_\mathrm{S}(t)/\partial t) \rangle$ is the power,
i.e., the time derivative of the work,
and
\begin{align}
\dot{Q}_\mathrm{S}^k(t) = \frac{i}{\hbar} \langle [ \hat{H}_\mathrm{I}^k(t), {\hat{H}_\mathrm{S}}(t) ] \rangle
\label{eq:SHC_def}
\end{align}
is the change in the system energy due to the coupling with the $k$th bath.
This is identical to the definition used in the QME approach,
in which the system Hamiltonian is identified as the internal energy.
The second expression for the heat current is derived through consideration of the rate of decrease of the bath energy,
$\dot{Q}_\mathrm{B}^k(t) \equiv - d\langle \hat{H}_\mathrm{B}^k(t) \rangle/dt$.
We call this current the "bath heat current" (BHC).
Using the Heisenberg equations, the BHC can be rewritten as
\begin{align}
\dot{Q}_\mathrm{B}^k(t)
= \dot{Q}_\mathrm{S}^k(t)
  + \frac{d}{dt}\langle \hat{H}_\mathrm{I}^k(t) \rangle
  + \sum_{k' \ne k} \dot{q}_{k,k'},
\label{eq:bath_heat}
\end{align}
where
\begin{align}
\dot{q}_{k,k'}(t)
= \frac{i}{\hbar} \langle [ \hat{H}_\mathrm{I}^k(t), \hat{H}_\mathrm{I}^{k'}(t) ] \rangle.
\label{eq:current_I_def}
\end{align}
The second term on the right hand side of Eq.(\ref{eq:bath_heat}) vanishes under steady-state conditions and in the limit of a weak system-bath coupling.
The third term contributes to the heat current even under steady-state conditions,
while it vanishes in the weak coupling limit.
The third term, which is of purely quantum mechanical origin, as can be seen from the definition,
is the main difference between the SHC and the BHC.
This term plays a significant role in the case that
the $k$th and $k'$th system-bath interactions are non-commuting and each system-bath coupling is strong.
We also note that because this third term is of greater than fourth-order in the system-bath interaction,
it does not appear in the second-order QME approach.
{Therefore only non-perturbative approaches \cite{Gelbwaser,Strasberg,Newman,Carrega,Esposito15PRL,Esposito15PRB,Schmidt} including higher-order QME approaches\cite{Wu} may allow us to reveal the features that we are going to discuss in the present study.}
Hereafter, we refer to this term as the "correlation among the system-bath interactions" (CASBI).
For a mesoscopic heat-transport system, including nanotubes and nanowires, each system component is coupled to a different bath
( i.e., each $\hat{V}_k$ acts on a different Hilbert space),
and for this reason, the CASBI contributions vanish.
By contrast, for a microscopic system, including single-molecular junctions and superconducting qubits, the CASBI contribution plays a significant role. Using our two definitions of the heat current, we are able to elucidate the important dynamic properties of microscopic systems and clearly demonstrate how their quantum mechanical nature is manifested.

\section{The First and Second Laws of Thermodynamics}\label{sec:thermodynamics}
We can obtain the first law of thermodynamics by summing Eq.(\ref{eq:bath_heat}) over all $k$:
\begin{align}
\sum_{k=1}^K \dot{Q}_\mathrm{B}^k(t)
= \frac{d}{dt} \langle \hat{H}_\mathrm{S}(t) + \sum_{k=1}^{K} \hat{H}_\mathrm{I}^k(t) \rangle
  - \dot{W}(t).
\label{eq:first_law}
\end{align}
The quantity $\hat{H}_\mathrm{S}(t) + \sum_{k=1}^K \hat{H}_\mathrm{I}^k (t)$ is identified as the internal energy,
because the contributions of $\dot{q}_{k,k'}$ cancel out.

The derivation of the second law of thermodynamics is presented in Appendix \ref{sec:second_law}.
In a steady state without external driving forces, the second law is expressed as
\begin{align}
- \sum_{k=1}^K \frac{\dot{Q}_\mathrm{B}^k}{T_k} \ge 0,
\end{align}
while with a periodic external driving force, it is given by
\begin{align}
- \sum_{k=1}^K \frac{Q_\mathrm{B}^{\mathrm{cyc},k}}{T_k} \ge 0,
\end{align}
where $Q_\mathrm{B}^{\mathrm{cyc},k} = \oint_\mathrm{cyc} dt\, \dot{Q}_\mathrm{B}^k(t)$ is the heat absorbed or released per cycle.
The second law without a driving force can be rewritten in terms of the SHC as
\begin{align}
- \sum_{k=1}^K \frac{Q_\mathrm{S}^{\mathrm{cyc},k} }{T_k}
\ge \sum_{k,k'=1}^K \frac{ q_{k,k'}^\mathrm{cyc} }{ T_k }.
\label{eq:second_law2}
\end{align}
When the right-hand side (rhs) of Eq.(\ref{eq:second_law2}) is negative,
the left-hand side (lhs) can also take negative values.
However, this contradicts the Clausius statement of the second law,
i.e., that heat never flows spontaneously from a cold body to a hot body.
As we show in Sec. \ref{sec:numerics},
it is necessary to include the $\dot{q}_{k,k'}$ terms to have a thermodynamically valid description.

\section{Reduced Description of Heat Currents}\label{sec:reduced}
In order to calculate the heat current from the reduced system dynamics,
we must evaluate the expectation value of the collective bath coordinate.
To do so, we adapt a generating functional approach \cite{Schwinger} by adding the source term, $f_k(t)$,
for the $k$th interaction Hamiltonian as
\begin{align}
\hat{V}_k \hat{X}_k \to \hat{V}_{k,f}(t) \hat{X}_k \equiv ( \hat{V}_k + f_k(t) ) \hat{X}_k.
\end{align}
{Here, in order to evaluate an expectation value, we add the source term to the ket (left) side of the density operator, which does not change a role of the system-bath interaction in the time-evolution operator.}
The interaction representation of any operator, $\hat A$, with respect to the non-interacting Hamiltonian,$\hat{H}_\mathrm{S}(t) + \sum_k \hat{H}_\mathrm{B}^k$ is expressed as $\tilde{A}(t)$. The total density operator, $\tilde{\rho}_\mathrm{tot}(t)$, with the source term is then denoted by $\tilde{\rho}_{\mathrm{tot},f}$.
This source term enables us to have a collective bath coordinate with the functional derivative as
\begin{align}
\tilde{X}_k(t) \tilde{\rho}_\mathrm{tot}(t)
= i \hbar \frac{\delta}{\delta f_k(t)} \left. \tilde{\rho}_{\mathrm{tot},f}(t) \right|_{f = 0},
\end{align}
Then, for example, the SHC and the interaction Hamiltonian are expressed in terms of the operators in the system space as
\begin{align}
\dot{Q}_\mathrm{S}^k(t)
= \mathrm{Tr_S}\left \{ \left[ \tilde{H}_\mathrm{S}(t), \tilde{V}_k \right]
  \frac{\delta}{\delta f_k(t)} \tilde{\rho}_f(t)|_{f=0} \right \}
\label{eq:SHC_functional}
\end{align}
and
\begin{align}
\left\langle \hat{H}_\mathrm{I}^k(t) \right\rangle
= i \hbar \, \mathrm{Tr_S} \left \{
  \tilde{V}_k \frac{\delta}{\delta f_k(t)} \tilde{\rho}_f(t)|_{f=0} \right \},
\label{eq:Hint_functional}
\end{align}
where $\tilde{\rho}(t) \equiv \mathrm{Tr_B}\{ \tilde{\rho}_\mathrm{tot}(t)\}$
is the reduced density operator of the system obtained by tracing out all bath degrees of freedom.
This is expressed in the form of a second-order cumulant expansion,
which is exact in the present case, due to the bosonic nature of the bath:
\begin{align}
\tilde{\rho}_f(t)
= \mathcal{T}_+ \left\{ \mathcal{U}_{\mathrm{IF},f}(t) \right\} \hat{\rho}(0).
\end{align}
Here, $\mathcal{U}_\mathrm{IF}(t) = \prod_{k=1}^K \exp[ \int_0^t ds\, W_k(s) ]$
is the Feynman-Vernon influence functional in operator form,
and $\mathcal{T}_+\{ \ldots \}$ is the time-ordering operator,
where the operators in $\{ \ldots \}$ are arranged in chronological order.
Here, the initial state is taken to be the product state of the reduced system and the bath density operators,
$\hat{\rho}_\mathrm{tot}(0) = \hat{\rho}(0)\prod_{k=1}^K \hat{\rho}_\mathrm{B}^{k,\mathrm{eq}}$,
where $\hat{\rho}_\mathrm{B}^{k,\mathrm{eq}}$ is the canonical density operator for the $k$th bath.
Note that, by generalizing the influence functional, \cite{Tanimura14, Tanimura15}
we can extend the present result to the case of a mixed initial state of the reduced system and the bath.
The operators of the influence phase are defined by
\begin{align}
W_k(s)
= &\, \int_0^s du\, \tilde{\Phi}_k(s)
      \left[ C_{k}^\mathrm{R}(s-u) \tilde{\Phi}_k(u) \right.
\notag \\
  &\, \left. - C_{k}^\mathrm{I}(s-u) \tilde{\Psi}_k(u) \right],
\label{eq:W_IF}
\end{align}
where we have introduced the two superoperators $\hat{\Phi}_k \hat{\rho} = (i/\hbar) [\hat{V}_k, \hat{\rho} ]$
and $\hat{\Psi}_k \hat{\rho} = (1/\hbar) \{ \hat{V}_k, \hat{\rho} \}$,
and $C_k^\mathrm{R}(s)$ and $C_k^\mathrm{I}(s)$ are the real and imaginary parts of $C_k(s)$.
Thus, by applying the functional derivative with respect to $f_k(t)$, to the reduced density operator,
we obtain the following relation:
\begin{align}
i \hbar \frac{\delta}{\delta f_k(t)} \tilde{\rho}_f(t)|_{f=0}
= &\, - \mathcal{T}_+\left\{ \int_0^t ds
      \left[ C_k^\mathrm{R}(t-s) \tilde{\Phi}_k(s) \right. \right.
\notag \\
  &\, \left.\left. - C_k^\mathrm{I}(t-s) \tilde{\Psi}_k(s) \right]
      \mathcal{U}_\mathrm{IF}(t) \right\} \hat{\rho}(0).
\label{eq:IF_equality}
\end{align}
From the generating functional, for the SHC, we obtain the expression
\begin{align}
\dot{Q}_\mathrm{S}^k(t)
= &\, \int_0^t ds \left( C_{k}^\mathrm{R}(t-s)
      \frac{i}{\hbar} \langle [ \hat{A}_k(t), \hat{V}_k(s) ] \rangle \right.
\notag \\
  &\, \left. - C_{k}^\mathrm{I}(t-s)
      \frac{1}{\hbar} \langle \{ \hat{A}_k(t), \hat{V}_k(s) \} \rangle \right),
\label{eq:current_h_exp}
\end{align}
where $\hat{A}_k(t) \equiv (i/\hbar)[ \hat{H}_\mathrm{S}(t), \hat{V}_k(t) ]$.
In order to obtain an explicit expression for the BHC given in Eq.\eqref{eq:SHC_functional},
we need to evaluate the expectation value of the interaction energy.
From the generating functional, this is obtained as
\begin{align}
\left\langle \hat{H}_\mathrm{I}^k(t) \right\rangle
= &\, - \int_0^t ds \left( \bar{C}_k^\mathrm{R}(t-s)
      \frac{i}{\hbar} \langle [ \hat{V}_k(t), \hat{V}_k(s) ] \right.
\notag \\
  &\, \left. - C_k^\mathrm{I}(t-s)
      \frac{1}{\hbar} \langle \{ \hat{V}_k(t), \hat{V}_k(s) \} \rangle \right).
 \label{eq:Hint_exp}
\end{align}
In order to obtain this expression,
we have divided the real part of the noise correlation function
into a short-time correlated part, expressed by a delta-function, and the remaining part, as
$C_{k}^R (s) = \bar{C}_{k}^R(s) + 2\Delta_k \delta(s)$.
The imaginary part of the noise correlation function is similarly divided into a delta-function part and the remaining part,
but in this case, we incorporate this delta-function part into the system Hamiltonian by renormalizing the frequency.
Taking the time derivative of Eq.(\ref{eq:Hint_exp}), we obtain the following:
\begin{align}
\frac{d}{dt} \left\langle \hat{H}_\mathrm{I}^k(t) \right\rangle
- \left \langle \frac{d\hat{V}_k(t)}{dt} \hat{X}_k(t) \right \rangle
= &\, - \int_0^t ds \left( \dot{\bar{C}}_k^\mathrm{R}(t-s)
      \frac{i}{\hbar} \langle [ \hat{V}_k(t), \hat{V}_k(s) ] \right.
\notag \\
  &\, \left. - \dot{C}_k^\mathrm{I}(t-s)
      \frac{1}{\hbar} \langle \{ \hat{V}_k(t), \hat{V}_k(s) \} \rangle \right)
\notag \\
  &\, + \frac{i}{\hbar} \Delta_k
      \left\langle \left[ \frac{d \hat{V}_k(t)}{dt}, \hat{V}_k(t) \right] \right\rangle
\notag \\
  &\, + \frac{2}{\hbar} C_{k}^\mathrm{I}(0) \langle \hat{V}_k^2(t) \rangle.
\label{eq:dHint_exp}
\end{align}
The lhs of the above equation is identical to $\dot{Q}_\mathrm{B}^k(t)$,
because we have
\begin{align}
\left\langle \frac{ d\hat{V}_k(t) }{dt} \hat{X}_k(t) \right\rangle
= - \dot{Q}_\mathrm{S}^k(t) - \sum_{k' \ne k} \dot{q}_{k,k'}(t),
\label{eq:dVdt_equality1}
\end{align}
which is derived from the Heisenberg equation of motion for $\hat{V}_k$,
\begin{align}
\frac{d\hat{V}_k(t)}{dt}
= \hat{A}_k(t)
  - \sum_{k' \ne k} \frac{i}{\hbar} [ \hat{V}_k(t), \hat{V}_{k'}(t) ] \hat{X}_{k'}(t).
\label{eq:V_Heisenberg}
\end{align}
Accordingly, using the relation
\begin{align}
\frac{i}{\hbar} \left\langle \left[ \frac{d \hat{V}_k(t)}{dt}, \hat{V}_k(t) \right] \right\rangle
= &\, \sum_{k' \ne k} \int_0^t ds \left( C_{k'}^\mathrm{R}(t-s)
      \frac{i}{\hbar} \left \langle [ \hat{B}_{k,k'}(t), \hat{V}_{k'}(s) ] \right\rangle \right.
\notag \\
  &\, \left. - C_{k'}^\mathrm{I}(t-s)
      \frac{1}{\hbar} \langle \{ \hat{B}_{k,k'}(t), \hat{V}_{k'}(s) \} \rangle \right)
\notag \\
  &\, + \frac{i}{\hbar} \langle [ \hat{A}_k(t), \hat{V}_k(t) ] \rangle,
\label{eq:dVdt_equality2}
\end{align}
where $\hat{B}_{k,k'} \equiv (i/\hbar)^2 [ [ \hat{V}_k, \hat{V}_{k'} ], \hat{V}_k ]$,
we obtain the following as the final expression for the BHC:
\begin{align}
\dot{Q}_\mathrm{B}^k(t)
= &\, - \int_0^t ds \left( \dot{\bar{C}}_k^\mathrm{R}(t-s)
      \frac{i}{\hbar} \langle [ \hat{V}_k(t), \hat{V}_k(s) ] \rangle \right.
\notag \\
  &\, \left. - \dot{C}_k^\mathrm{I}(t-s)
      \frac{1}{\hbar} \langle \{ \hat{V}_k(t), \hat{V}_k(s) \} \rangle \right)
\notag \\
  &\, + \frac{2}{\hbar} C_k^\mathrm{I}(0) \langle \hat{V}_k^2(t) \rangle
      + \frac{i}{\hbar} \Delta_k \langle [ \hat{A}_k(t), \hat{V}_k(t) ] \rangle
\notag \\
  &\, + \Delta_k \sum_{k' \ne k} \int_0^t ds \left( C_{k'}^\mathrm{R}(t-s)
      \frac{i}{\hbar} \langle [ \hat{B}_{k,k'}(t), \hat{V}_{k'}(s) ] \rangle \right.
\notag \\
  &\, \left. - C_{k'}^\mathrm{I}(t-s)
      \frac{1}{\hbar} \langle \{ \hat{B}_{k,k'}(t), \hat{V}_{k'}(s) \} \rangle \right).
\label{eq:dQ_final}
\end{align}

Now that we have obtained the explicit expressions for the SHC and BHC,
the remaining task is to evaluate these expressions in a numerically rigorous manner.
This was carried out using the HEOM approach.

\section{Hierarchal Equations of Motion Approach}\label{sec:heom}
When the noise correlation function, Eq. \eqref{eq:noise},
is written as a linear combination of exponential functions and a delta function,
$C_k(t) = \sum_{j=0}^{J_k} ( c'_{k_j} + i c''_{k_j}) e^{-\gamma_{k_j}|t|} + 2\Delta_k \delta(t)$,
which is realized for the Drude, \cite{Tanimura88,Ishizaki05,Tanimura06,Tanimura14, Tanimura15, Kato15}
Lorentz, \cite{KramerFMO,Nori12}
and Brownian \cite{TanakaJPSJ09, YanBO12, Liu} cases (and combinations thereof \cite{TanimruaJCP12}),
we can obtain the reduced equations of motion as the HEOM.
The HEOM consist of the following set of equations of motion for the auxiliary density operators (ADOs):
\begin{align}
\hat{\rho}_{\vec{n}_1, \ldots, \vec{n}_K}(t)
\equiv &\, \mathcal{T}_+\left\{
           \exp\left[ - \frac{i}{\hbar} \int_0^t ds\, \mathcal{L}(s) \right] \right\}
\notag \\
       &\, \times \mathcal{T}_+\left\{ \prod_{k=1}^K \prod_{j=0}^{J_k}
           \left[ - \int_0^t ds\, e^{-\gamma_{k_j}(t-s)} \tilde{\Theta}_{k_j}(s) \right]^{n_{k_j}}
           \mathcal{U}_\mathrm{IF}(t) \right\}
\notag \\
       &\, \times \hat{\rho}(0).
\label{eq:ADO}
\end{align}
Here, we have $\hat{\Theta}_{k_j} \equiv c'_{k_j} \hat{\Phi}_k - c''_{k_j} \hat{\Psi}_k$
and $\mathcal{L}(t) \hat{\rho} = [ \hat{H}_\mathrm{S}(t), \hat{\rho} ]$.
Each ADO is specified by the index $\vec{n}_k = ( n_{k_0}, \ldots, n_{k_{J_k}})$ with $k = 1, \ldots, K$,
where each element takes an integer value larger than zero.
The ADO for which all elements are zero, $\vec{n}_0 = \cdots = \vec{n}_K = 0$,
corresponds to the actual reduced density operator.
Taking the time derivative of Eq.(\ref{eq:ADO}),
the equations of motion for the ADOs are obtained as
\begin{align}
\frac{\partial}{\partial t} \hat{\rho}_{\vec{n}_1, \ldots, \vec{n}_K}(t)
= &\, - \left[ \frac{i}{\hbar} \mathcal{L}(t)
      + \sum_{k=1}^K \sum_{j=0}^{J_k} n_{k_j} \gamma_{k_j} \right]
      \hat{\rho}_{\vec{n}_1, \ldots, \vec{n}_K}(t)
\notag \\
  &\, - \sum_{k=1}^K \Delta_k \hat{\Phi}_k^2
      \hat{\rho}_{\vec{n}_1, \ldots, \vec{n}_K}(t)
\notag \\
  &\, - \sum_{k=1}^K \hat{\Phi}_k \sum_{j=0}^{J_k}
      \hat{\rho}_{\vec{n}_1, \ldots, \vec{n}_k + \vec{e}_j, \ldots, \vec{n}_K}(t)
\notag \\
  &\, - \sum_{k=1}^K \sum_{j=0}^{J_k} n_{k_j} \hat{\Theta}_{k_j}
      \hat{\rho}_{\vec{n}_1, \ldots, \vec{n}_k - \vec{e}_j, \ldots, \vec{n}_K}(t),
\label{eq:HEOM}
\end{align}
where $\vec{e}_j$ is the unit vector along the $j$th direction.
The HEOM consist of an infinite number of equations,
but they can be truncated at finite order by ignoring all $k_j$ beyond the value
at which $\sum_{k_j} n_{k_j}$ first exceeds some appropriately large value $N$.

Employing the noise decomposition of the HEOM approach
for the noise correlation functions in Eqs. (\ref{eq:current_h_exp}) and (\ref{eq:dQ_final}),
and comparing the resulting expressions with the definition of the ADOs given in Eq.(\ref{eq:ADO}),
we can evaluate the SHC and BHC in terms of the ADOs as
\begin{align}
\dot{Q}_\mathrm{S}^k(t)
= &\, -\sum_{j=0}^{J_k} \mathrm{Tr}\{
      \hat{A}_k(t) \hat{\rho}_{\vec{0},\ldots,\vec{e}_j,\ldots,\vec{0}}(t) \}
\notag \\
  &\, + \Delta_k \frac{i}{\hbar}
      \mathrm{Tr}\{ [ \hat{A}_k(t), \hat{V}_k ] \hat{\rho}(t) \}
\end{align}
and
\begin{align}
\dot{Q}_\mathrm{B}^k(t)
= &\, - \sum_{j=0}^{J_k} \gamma_{k_j} \mathrm{Tr}\{
      \hat{V}_k \hat{\rho}_{\vec{0}, \ldots, \vec{e}_j, \ldots, \vec{0}}(t) \}
\notag \\
  &\, + \frac{2}{\hbar} C_{k}^\mathrm{I}(0)
      \mathrm{Tr}\{ \hat{V}_k^2 \hat{\rho}(t) \}
      + \frac{i}{\hbar} \Delta_k
      \mathrm{Tr}\{ [ \hat{A}_k(t), \hat{V}_k ] \hat{\rho}(t) \}
\notag \\
  &\, + \Delta_k \sum_{k' \ne k} \sum_{j=0}^{J_{k'}} \mathrm{Tr}\{
      \hat{B}_{k,k'} \hat{\rho}_{\vec{0}, \ldots, \vec{e}_j, \ldots, \vec{0}}(t) \}
\notag \\
  &\, + \Delta_k \sum_{k' \ne k } \frac{i}{\hbar} \Delta_{k'}
      \mathrm{Tr}\{ [ \hat{B}_{k,k'}, \hat{V}_{k'} ] \hat{\rho}(t) \}.
\end{align}
It is important to note that the steady-state solution of the HEOM is an entangled state of the system and the baths;
for example, for a static system coupled to a single bath ($k=1$),
the steady-state solution of the HEOM takes the form
$\hat{\rho} \propto \mathrm{Tr_B}\{ \exp[-\beta(\hat{H}_\mathrm{S}+\hat{H}_\mathrm{I}^1+\hat{H}_\mathrm{B}^1 ]\}$.
\cite{Tanimura14,Tanimura15}

\section{Numerical Illustration}\label{sec:numerics}
\begin{figure}
\centering
\includegraphics[width=0.4\textwidth]{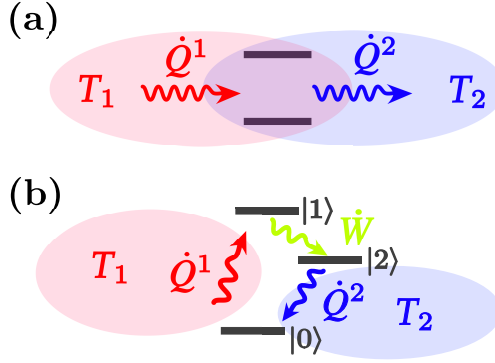}
\caption{ Schematic depiction of (a) the non-equilibrium spin-boson model and (b) the three-level heat engine model.}
\label{fig:model}
\end{figure}

To demonstrate the role of the CASBI in the heat current,
we carried out numerical simulations for a non-equilibrium spin-boson model \cite{Segal05,Thoss,Ruokola,Saito,Wang}
and a three-level heat engine model \cite{Scovil,Geva,Correa,Xu} with the HEOM approach (Fig. \ref{fig:model}).
{Here, we focus on investigating the steady-state heat currents, which are computed from Eq.(\ref{eq:HEOM}) using the Runge-Kutta method to numerically integrate them until convergence to the steady state is realized.}
We assume that the spectral density of each bath takes the Drude form,
$J_k(\omega) = \eta_k \gamma^2 \omega/( \omega^2 + \gamma^2 )$,
where $\eta_k$ is the system-bath coupling strength, and $\gamma$ is the cutoff frequency.
A Pad{\'e} spectral decomposition scheme is employed to obtain the expansion coefficients of the noise correlation functions. \cite{YanPade10A,YanPade10B,Hu}
{
The accuracy of numerical results is checked by increasing the values of $J_1, \ldots, J_K$ and $N$ until convergence is reached.}

%

\subsection{Non-equilibrium spin-Boson model}
The non-equilibrium spin-boson model studied here consists of a two-level system
coupled to two bosonic baths at different temperatures.
This model has been employed extensively as the simplest heat-transport model.
The system Hamiltonian is given by $\hat{H}_\mathrm{S} = ( \hbar \omega_0/2 ) \sigma_z$.
We consider the case in which the system is coupled to the first bath through $\hat{V}_1 = \sigma_x$
and to the second bath through $\sigma_x$ and $\sigma_z$ in the form
$\hat{V}_2 = ( s_x \sigma_x + s_z \sigma_z)/( s_x^2 + s_z^2 )^{1/2}$.
In order to investigate the difference between the SHC and BHC,
we consider the case $s_z \ne 0$,
because otherwise the CASBI term vanishes,
and thus the SHC coincides with the BHC.
(This is the case that most of previous investigations have considered.)
It should be noted that the CASBI has a purely quantum origin, as can be seen from its definition.
We chose $T_1 = 2.0 \hbar\omega_0/k_B, T_2 = 1.0\hbar\omega_0/k_B, \gamma = 2.0\omega_0$ and $s_z = 1$.

\begin{figure}
\centering
\includegraphics[width=0.4\textwidth]{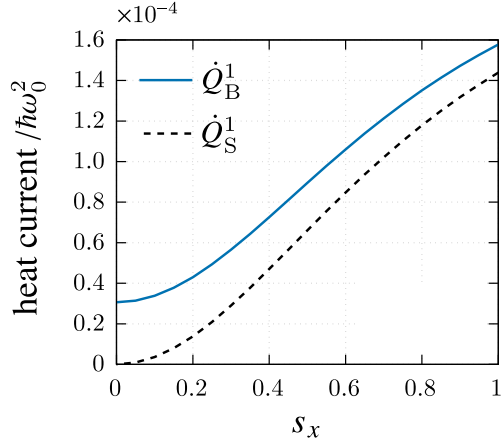}
\caption{The heat currents of the non-equilibrium spin-boson model are plotted
as functions of $s_x$ in order to illustrate the effect of the fact that $V_1$ and $V_2$ do not commute.
The solid (blue) and the dashed (black) curves represent $\dot{Q}_\mathrm{B}^1$
and $\dot{Q}_\mathrm{S}^1$, which correspond to the BHC and SHC.}
\label{fig:nesb_sx}
\end{figure}

Figure \ref{fig:nesb_sx} depicts the role of non-commuting component of the $V_1$ and $V_2$ interactions
in the SHC and BHC processes in the steady state as functions of $s_x$.
Here, the system-bath coupling strengths are set to $\eta_1 = \eta_2 = 0.01\omega_0$.
Even when the system Hamiltonian commutes with the second interaction Hamiltonian in the case $s_x = 0$
(i.e., even when the system couples to the second bath in a non-dissipative manner with the interaction $s_z \sigma_z$
as $[ \hat{H}_\mathrm{S}, \hat{H}_\mathrm{I}^2]=0$),
non-zero heat current arises due to the CASBI contribution, $\dot{q}_{1,2}$.
This is because the Hamiltonian of the system plus system-bath interactions does not commute with the second interaction
(i.e., $[ \hat{H}_\mathrm{S} + \sum_{k=1,2} \hat{H}_\mathrm{I}^k, \hat{H}_\mathrm{I}^2 ] = [ \hat{H}_\mathrm{I}^1, \hat{H}_\mathrm{I}^2 ] \ne 0$),
while the system Hamiltonian and the second interaction Hamiltonian do commute.

\begin{figure}
\centering
\includegraphics[width=0.4\textwidth]{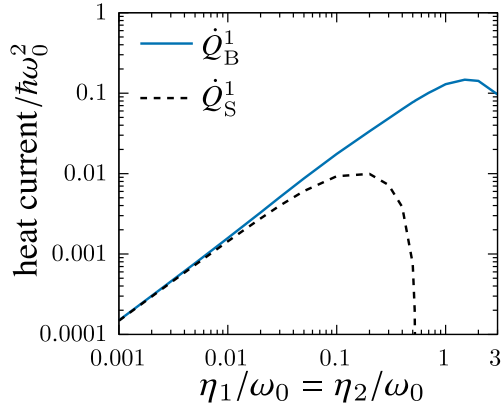}
\caption{The SHC and BHC corresponding to $\dot{Q}_\mathrm{S}^1$ (black dashed curve) and $\dot{Q}_\mathrm{B}^1$ (blue solid curve)
for the non-equilibrium spin-boson model as functions of the system-bath coupling.}
\label{fig:nesb_eta}
\end{figure}

Figure \ref{fig:nesb_eta} depicts the heat currents
as functions of the system-bath coupling strength for the case $s_x = s_z = 1$.
In the weak system-bath coupling regime,
the SHC and BHC increase linearly with the coupling strength in similar manners.
It is thus seem that in this case, the CASBI contribution is minor.
As the strength of the system-bath coupling increases,
the difference between the SHC and BHC becomes large:
While $\dot Q_S^1$ decreases after reaching a maximum value near $\eta_1 = \eta_2 = 0.2\omega_0$,
the CASBI contribution, $\dot{q}_{1,2}$, dominates the BHC,
and as a result, it remains relatively large.
Thus, in this regime, the SHC becomes much smaller than the BHC.
In the very strong coupling regime,
the SHC eventually becomes negative,
which indicates the violation of the second law.
In order to eliminate such non-physical behavior, we have to include the $\dot{q}_{1,2}$ term in the definition of the SHC.
Note that the differences between the SHC and BHC described above vanish for $s_z = 0$,
and hence in this case, there is no negative current problem.
This is the case considered in most previous investigations.

%

\subsection{Three-level heat engine model}
\begin{figure}
\centering
\includegraphics[width=0.4\textwidth]{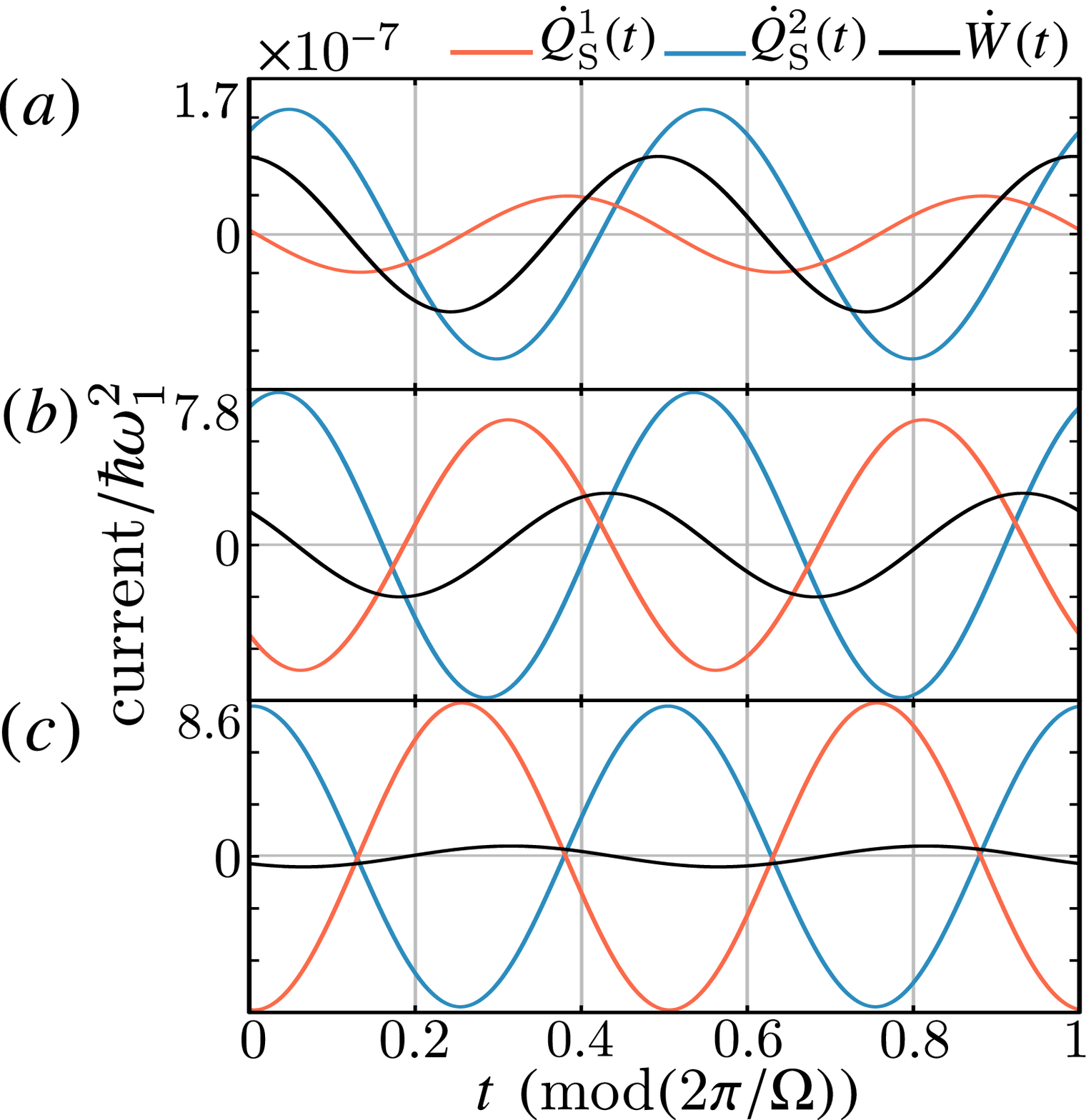}
\caption{The SHC of the first bath, $\dot{Q}_\mathrm{S}^1$ (red curve),
that of the second bath, $\dot{Q}_\mathrm{S}^2$ (blue curve),
and the power, $\dot{W}$ (black curve), are plotted as functions of time for (a) weak ($\eta_1 = 0.01\omega_1$),
(b) intermediate ($\eta_1 = 0.10\omega_1$),
and (c) strong ($\eta_1 = 1.00\omega_1$) coupling to the first bath
with fixed weak coupling to the second bath ($\eta_2 = 0.001\omega_1$).
The temperatures of the first and second baths are $T_1=10\hbar\omega_1/k_\mathrm{B}$ and $T_2=\hbar\omega_1/k_\mathrm{B}$.
The time period of $1$ corresponds to one cycle of the external force.
Each curve is properly adjusted as
$\dot{Y} \to \dot{Y} - \frac{1}{2}( \max_t\{ \dot{Y}(t) \} + \min_t\{ \dot{Y}(t) \})$
for $Y = Q_\mathrm{S}^1, Q_\mathrm{S}^2$, and $W$. }
\label{fig:three_time}
\end{figure}

The three-level heat engine model considered here consists of three states,
denoted by $| 0 \rangle$, $| 1 \rangle$, and $| 2 \rangle$, coupled to two bosonic baths.
The system is driven by a periodic external field with frequency $\Omega$.
The system Hamiltonian is expressed as
\begin{align}
\hat{H}_\mathrm{S}(t)
= \sum_{i=0,1,2} \hbar \omega_i | i \rangle \langle i | + g ( e^{-i\Omega t} | 1 \rangle \langle 2 | + \mathrm{H.c.})
\end{align}
with $\omega_1 > \omega_2 > \omega_0$.
The system-bath interactions are defined as
$\hat{V}_1 = | 0 \rangle \langle 1| + | 1 \rangle \langle 0|$
and $\hat{V}_2 = | 0 \rangle \langle 2 | + | 2 \rangle \langle 0 |$.
We set $\omega_0 = 0$ without loss of generality.
Roughly stated, this model acts as a quantum heat engine
when population inversion between the two excited states, $|1\rangle$ and $|2\rangle$, occurs.
This can be realized in the case that
the temperature of the first bath, $T_1$, is sufficiently higher than that of the second bath, $T_2$.
Using this model, we analyze the work and the heat per cycle,
i.e., $Y^{\mathrm{cyc}} = \lim_{ t \to \infty } \int_t^{ t + 2\pi/\Omega } \dot{Y}(t') dt'$
for $Y = W, Q_\mathrm{S}^{k}$, and $Q_\mathrm{B}^{k}$ with $k=1$ or $2$.
We set $\omega_2 = 0.5\omega_1$, $\Omega = 0.5\omega_1, \gamma = 2.0\omega_1$, and $g = 0.1\hbar\omega_1$.

In Figure \ref{fig:three_time}, we illustrate the time dependences of the two SHC,
$\dot{Q}_\mathrm{S}^1$ and $\dot{Q}_\mathrm{S}^2$, and the power, $W$,
that arise from transitions between $|0 \rangle$ and $|1 \rangle$, $|0 \rangle$ and $|2 \rangle$, and $|1 \rangle$ and $|2 \rangle$, respectively,
for several values of the first bath coupling,
with the second bath coupling fixed ($\eta_2 = 0.001\omega_1$).
The figure depicts $\dot{Q}_\mathrm{S}^1, \dot{Q}_\mathrm{S}^2$, and $\dot{W}$ for one cycle of the external force.
We set $t=0$ and $t=1$ to correspond to the maxima of the cyclic driving field.
The time delays observed for $\dot{Q}_\mathrm{S}^1$, $\dot{Q}_\mathrm{S}^2$, and $\dot{W}$
imply that the transition $|0\rangle \to |1\rangle \to |2\rangle$ is cyclic.
This behavior can be regarded as a microscopic manifestation of a quantum heat engine.
The periods of the currents and power are, however, half as long as the period of the driving field.
This is because,
while the transition for the work production is induced by an even number of system-field interactions,
the second-order interaction that involves components with frequency $2\Omega$,
which is twice that of the system-field interaction, becomes dominant.

The phases of $\dot{Q}_\mathrm{S}^1$, $\dot{Q}_\mathrm{S}^2$ and $W$ depend on the first bath coupling.
Because the strength of the second bath coupling and the external fields are weak,
all of these changes are a consequence of the change in $\dot{Q}_\mathrm{S}^1$.
When the first bath coupling is weak,
the first SHC, $\dot{Q}_\mathrm{S}^1$, which is the current from the high-temperature heat bath,
cannot follow the change of the external field and hence exhibits a delay in its response to the decrease of the heat current
that occurs at the maxima of the field intensity.
As a result, the power, $W$, and the heat current for the low-temperature bath, $\dot{Q}_\mathrm{S}^2$,
exhibit successively delayed response.
When the first bath coupling is strong,
$\dot{Q}_\mathrm{S}^1$ closely follows the variation of the field.
The time delays of $\dot{Q}_\mathrm{S}^2$ and $\dot{W}$ also decrease as first bath coupling increases.

\begin{figure}
\centering
\includegraphics[width=0.4\textwidth]{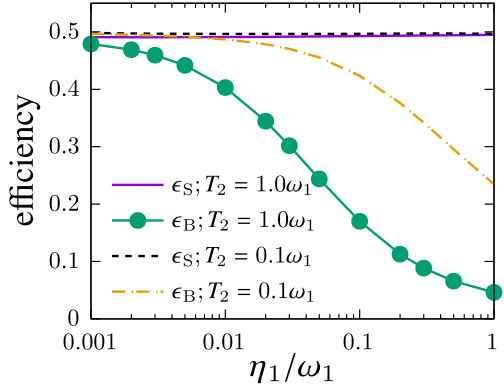}
\caption{The efficiencies of the three-level heat engine
calculated as functions of the coupling to the first bath, $\eta_1$,
with fixed weak coupling to the second bath ($\eta_2 = 0.001\omega_1$).
Here, we consider $\epsilon_\mathrm{S}=-W^\mathrm{cyc}/Q_\mathrm{S}^{\mathrm{cyc},1}$ (solid line)
and $\epsilon_\mathrm{B}=-W^\mathrm{cyc}/Q_\mathrm{B}^{\mathrm{cyc},1}$ (curve with circles) in the high temperature case,
with $T_2 = 1.0\hbar\omega_1/k_\mathrm{B}$,
and $\epsilon_\mathrm{S}$ (dashed curve) and $\epsilon_\mathrm{B}$ (dash-dotted line)
in the low temperature case, with $T_2 = 0.1\hbar\omega_1/k_\mathrm{B}$.}
\label{fig:three_eff}
\end{figure}

In Fig. \ref{fig:three_eff},
we depict the system efficiency, $\epsilon_\mathrm{S} \equiv -W^\mathrm{cyc}/Q_\mathrm{S}^{\mathrm{cyc},1}$,
and the bath efficiency, $\epsilon_\mathrm{B} \equiv -W^\mathrm{cyc}/Q_\mathrm{B}^{\mathrm{cyc},1}$,
as functions of the strength of the coupling to the first bath
with fixed strength of the coupling to the second bath, $\eta_2=0.001\omega_1$. Here, $W^{\mathrm{cyc}}$ and $Q^{\mathrm{cyc}}$ represent the time average of $W$ and $Q$ per cycle. We consider a high temperature case with $T_2=\hbar\omega_1/k_\mathrm{B}$
and a low temperature case with $T_2=0.1\hbar\omega_1/k_\mathrm{B}$,
with the fixed ratio $T_2 / T_1 = 0.1$.

\begin{figure}
\centering
\includegraphics[width=0.4\textwidth]{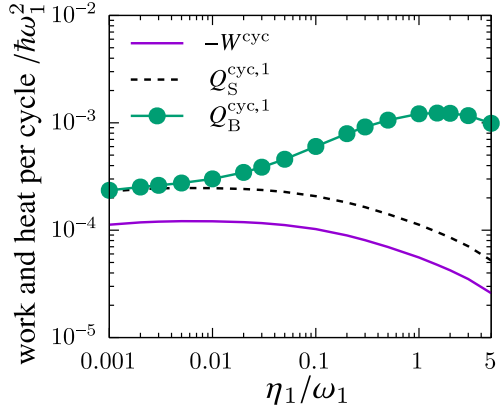}
\caption{The work and heat per cycle of the three-level heat engine
calculated as functions of the coupling to the first bath, $\eta_1$,
with fixed weak coupling to the second bath ($\eta_2 = 0.001\omega_1$).
Here, we consider only the high temperature case, with $T_2 = 1.0\omega_1$.
The solid, dashed, and circle curves represent the work,
the system heat, $Q_\mathrm{S}^{\mathrm{cyc},1}$,
and the bath heat, $Q_\mathrm{B}^{\mathrm{cyc},1}$, respectively.}
\label{fig:three_work}
\end{figure}

While the system efficiency is weakly dependent on the strength of the first bath coupling,
regardless of the temperature,
the bath efficiency decreases as the strength of the first bath coupling increases,
in particular in the low temperature case.
The reason for this can be understood from Fig. \ref{fig:three_work}.
There, it is seen that $Q_\mathrm{S}^{\mathrm{cyc},1}$ decreases as the strength of the first bath coupling increases,
as a result of the strong suppression of the thermal activation by dissipation.
{The overall profiles of the work, $W^{\mathrm{cyc}}$ as well as $Q_\mathrm{S}^{\mathrm{cyc},2}$ (not shown) are similar to $Q_\mathrm{S}^{\mathrm{cyc},1}$. The $\eta_1$ dependence of $Q_\mathrm{S}^{\mathrm{cyc},1}$ and $W^{\mathrm{cyc}}$ follows that of $Q_\mathrm{S}^{\mathrm{cyc},2}$, because, under the present weak system and second bath interaction, the heat flow and work are determined by the capability of the second bath to drain the heat.\cite{Correa} }
Because we set the strength of the second bath coupling and the external fields are weak,
the work $-W^\mathrm{cyc}$ tends to follow the behavior of $Q_\mathrm{S}^{\mathrm{cyc},1}$,
as illustrated in Fig. \ref{fig:three_time},
whereas $Q_\mathrm{B}^{\mathrm{cyc},1}$ increases as the strength of the coupling to the first bath increases,
due to the CASBI contribution, $\dot{q}_{1,2}$.
As a result, the system efficiency, $-W^\mathrm{cyc}/Q_\mathrm{S}^{\mathrm{cyc},1}$,
does not change significantly,
whereas the bath efficiency, $-W^\mathrm{cyc}/Q_\mathrm{B}^{\mathrm{cyc},1}$,
decreases as a function of the first coupling strength.

In the strong coupling regime,
the bath efficiency in the low temperature case is larger than that in the high temperature case,
as depicted in Fig. \ref{fig:three_eff},
because as the temperature decreases,
the system coupled to the low temperature bath becomes less activated.
Note that if the strength of the coupling to the second bath is sufficiently large
that the system is in the non-perturbative regime,
the system efficiency decreases as the strength of the coupling to the first bath increases,
because in this case, a part of $-W$ goes to $Q_\mathrm{S}^{\mathrm{cyc},2}$.

Unlike in the non-equilibrium spin-boson case, the system efficiency is physically meaningful.
We believe, however, that the bath efficiency is more appropriate as the rigorous definition of the heat efficiency,
because the system efficiency does not include the contribution to the energy from the system-bath interactions, which must be regarded as a part of the system. The decrease of efficiency can be regarded as a quantum effect, because it originates from the discretization of the energy eigenstates.

%

\section{Concluding Remarks}\label{sec:conclusion}
In this paper,
we introduced an explicit expression for the bath heat current (BHC),
which includes contributions from the correlations among the system-bath interactions (CASBI).
The BHC reduces to the widely used system heat current (SHC) derived in terms of the system energy
under conditions of a weak system-bath coupling
or in the case that all system-bath interactions commute.
Our definition of the BHC can be applied to any system with any driving force
and any strength of the system-bath coupling.
We numerically examined the role of the CASBI using the HEOM approach.
We demonstrated this approach in the case of a non-equilibrium spin-boson system in which the CASBI contribution is necessary
to maintain consistency with thermodynamic laws in the strong system-bath coupling regime.
In the three-level heat-engine model,
we observed cyclic time evolution of the high-temperature heat current, the work, and the low-temperature heat current,
as in a classical heat engine.
When the system-bath coupling is weak,
there is a time delay between the variation of the external field
and the heat current of the high-temperature bath,
because this bath cannot follow variations of the system, due to the weakness of the system-bath coupling.
Contrastingly the heat current does not exhibit any time delay in the strong system-bath coupling case.
The efficiency defined using the BHC,
which is regarded as physically more appropriate than that defined using the SHC, decrease as the strength of the system-bath coupling increases.

{Although the definition of the heat current under non-steady-state condition is not clear,\cite{Esposito15PRB}} we can also apply our formulation to analysis of transient behavior,
in which the variation of the bath energy in time is experimentally measurable.
Because the HEOM approach is capable of calculating various physical quantities in non-equilibrium situations,
it would also be interesting to extend the present investigation to other quantum transport problems \cite{Ye}
by calculating higher-order cumulants \cite{Cerrillo}
and non-linear optical signals \cite{Agarwalla,Gao1,Gao2} to reveal the detailed physical properties of the dynamics.
We leave such problems to future studies.

\acknowledgments
Financial support from a Grant-in-Aid for Scientific Research (A26248005) from the Japan Society for the Promotion of Science is acknowledged.

\appendix
\section{Derivation of the second law}\label{sec:second_law}
In this Appendix,
we derive the Clausius inequality as the second law of thermodynamics for quantum steady states
by extending the result of Deffner and Jarzynski \cite{Deffner} for the classical case to the quantum case.
Note that this derivation is not limited to the case of reduced dynamics.
Because extension to the case of multiple heat baths is straightforward,
here we consider the case of a single bath, with the Hamiltonian
$\hat{H}(t) = \hat{H}_\mathrm{S}(t) + \hat{H}_\mathrm{I} + \hat{H}_\mathrm{B}$.
We consider the von Neumann entropy $\mathcal{H}(t)$ defined as
\begin{align}
\mathcal{H}(t) \equiv - \mathrm{Tr}\{ \hat{\rho}(t) \ln \hat{\rho}(t) \},
\end{align}
where $\hat{\rho}(t)$ is the total density operator.
The reduced density operator for the system is $\hat{\rho}_\mathrm{S}(t) \equiv \mathrm{Tr_B}\{ \hat{\rho}(t) \}$, while that for the bath is $\hat{\rho}_\mathrm{B}(t) \equiv \mathrm{Tr_S}\{ \hat{\rho}(t) \}$.
Then we introduce the von Neumann entropies of the system and bath as
$\mathcal{H}_\mathrm{S}(t) = - \mathrm{Tr}\{ \hat{\rho}_\mathrm{S}(t) \ln \hat{\rho}_\mathrm{S}(t) \}$
and $\mathcal{H}_\mathrm{B}(t) = - \mathrm{Tr}\{ \hat{\rho}_\mathrm{B}(t) \ln \hat{\rho}_\mathrm{B}(t) \}$, respectively.
Without loss of generality,
we assume that the total system is initially in the factorized state
$\hat{\rho}(0) = \hat{\rho}_\mathrm{S}(0) \hat{\rho}_\mathrm{B}^\mathrm{eq}$,
where $\hat{\rho}_\mathrm{B}^\mathrm{eq} = e^{-\beta \hat{H}_\mathrm{B}}/Z_B$
is the equilibrium density operator of the bath,
in which $Z_\mathrm{B}$ is the partition function of the bath.
In this case, the total entropy is merely the sum of the system and bath entropies:
\begin{align}
\mathcal{H}(0) = \mathcal{H}_\mathrm{S}(0) + \mathcal{H}_\mathrm{B}(0).
\end{align}
Because the von Neumann entropy is invariant under unitary evolution
(i.e. $\mathcal{H}(t) = \mathcal{H}(0)$),
and because the von Neumann entropy is sub-additive
(i.e. $\mathcal{H}(t) \le \mathcal{H}_\mathrm{S}(t) + \mathcal{H}_\mathrm{B}(t)$),
we have the following inequality for
$\Delta \mathcal{H}_\mathrm{S}(t) \equiv \mathcal{H}_\mathrm{S}(t) - \mathcal{H}_\mathrm{S}(0)$
and $\Delta \mathcal{H}_\mathrm{B}(t) \equiv \mathcal{H}_\mathrm{B}(t) - \mathcal{H}_\mathrm{B}(0)$ :
\begin{align}
\Delta \mathcal{H}_\mathrm{S}(t) + \Delta \mathcal{H}_\mathrm{B}(t) \ge 0.
\label{eq:inequality}
\end{align}
We now rewrite the von Neumann entropy of the bath as
\begin{align}
\mathcal{H}_\mathrm{B}(t)
& = - \mathrm{Tr}\{ \hat{\rho}_\mathrm{B}(t) \ln \hat{\rho}_\mathrm{B}^\mathrm{eq} \}
    - \left( \mathrm{Tr}\{ \hat{\rho}_\mathrm{B}(t) \ln \hat{\rho}_\mathrm{B}(t) \}
    - \mathrm{Tr}\{ \hat{\rho}_\mathrm{B}(t) \ln \hat{\rho}_\mathrm{B}^\mathrm{eq} \} \right)
\notag \\
& = \beta E_\mathrm{B}(t) - \beta F_\mathrm{B}
    - D( \hat{\rho}_\mathrm{B}(t) || \hat{\rho}_\mathrm{B}^\mathrm{eq} ),
\end{align}
where $E_\mathrm{B}(t) = \mathrm{Tr}\{ \hat{H}_\mathrm{B} \hat{\rho}_\mathrm{B}(t) \},
F_\mathrm{B} = - \beta^{-1} \ln Z_\mathrm{B}$,
and $D(\hat{\rho}||\hat{\sigma}) = \mathrm{Tr}\{ \hat{\rho} \ln \hat{\rho} \} - \mathrm{Tr}\{ \hat{\rho} \ln \hat{\sigma} \} \ge 0$
are the bath energy, the bath free energy, and the quantum relative entropy, respectively.
This leads to the condition
\begin{align}
\Delta \mathcal{H}_\mathrm{B}(t)
= \beta \Delta E_\mathrm{B}(t)
  - D(\hat{\rho}_\mathrm{B}(t)||\hat{\rho}_\mathrm{B}^\mathrm{eq})
\le \beta \Delta E_\mathrm{B}(t).
\end{align}
Then, using Eq.(\ref{eq:inequality}),
we obtain the inequality
\begin{align}
\Delta \mathcal{H}_\mathrm{S}(t) + \beta \Delta E_\mathrm{B}(t) \ge 0.
\label{eq:inequality2}
\end{align}

A quantum heat machine is subject to a periodic perturbation,
$\hat{H}_\mathrm{S}(t) = \hat{H}_\mathrm{S}(t+T)$,
where $T$ is the period of the perturbation.
After a sufficiently long time,
the system relaxes to a periodic steady state,
with $\hat{\rho}_\mathrm{S}(t) = \hat{\rho}_\mathrm{S}(t+T)$.
We separate the time of observation into a transient part (from $t=0$ to $n_0 T$) and a steady part (from $t=n_0 T$ to $n T$),
where $n_0$ is an integer sufficiently large to ensure that the system is in the periodic steady state at $n_0 T$.
The change in a variable $A$ from time $n_0 T$ to $n T$ is written $\Delta A^{n_0 \to n}=A(nT) - A(n_0T)$.
The inequality Eq.(\ref{eq:inequality2}) is then partitioned as
\begin{align}
\Delta \mathcal{H}_\mathrm{S}^{0 \to n_0}
+ \Delta \mathcal{H}_\mathrm{S}^{n_0 \to n}
+ \beta \Delta E_\mathrm{B}^{0 \to n_0}
+ \beta \Delta E_\mathrm{B}^{n_0 \to n}
\ge 0.
\end{align}
The second term on the right-hand side of this equation vanishes,
because the system is in a periodic steady state during the time from $n_0 T$ to $n T$.
The bath releases or absorbs a constant amount of heat per cycle,
\begin{align}
\Delta E_\mathrm{B}^{n_0 \to n} = (n - n_0) \Delta E_\mathrm{B}^\mathrm{cyc},
\end{align}
where $\Delta E_\mathrm{B}^\mathrm{cyc}$ is the change in the bath energy per cycle.
This leads to the inequality
\begin{align}
\beta \Delta E_\mathrm{B}^\mathrm{cyc} \ge 0
\end{align}
for large $n$.
Because the change of the bath energy is identical to the bath heat, $-Q_\mathrm{B}^\mathrm{cyc}$, the Clausius inequality is obtained.
With the straightforward extension of this derivation the case of multiple baths,
we obtain the general inequality
\begin{align}
- \sum_k \frac{ Q_\mathrm{B}^{\mathrm{cyc},k}}{T_k} \ge 0.
\end{align}
This is the result presented in Sec. \ref{sec:thermodynamics}.

\bibliography{aipsamp}

\begin{thebibliography}{99}
\bibitem{Ritort} F. Ritort, Adv. Chem. Phys. \textbf{137}, 31 (2008).
\bibitem{Seifert} U. Seifert, Rep. Prog. Phys. \textbf{75}, 126001 (2012).
\bibitem{Campisi} M. Campisi, P. H{\"a}nggi, and P. Talkner, Rev. Mod. Phys. \textbf{83}, 771 (2011).
\bibitem{Brandao} F. Brand{\~a}o, M. Horodecki, N. Ng, J. Oppenheim, and S. Wehner, Proc. Nat. Acad. Sci. \textbf{112}, 3275 (2013).
\bibitem{Trotzky} S. Trotzky, Y-A. Chen, A. Flesch, I. P. McCulloch, U. Schollw{\"o}ck, J. Eisert, and I. Bloch, Nat. Phys. \textbf{8}, 325 (2012).
\bibitem{Gemmer} J. Gemmer, M. Michel, and G. Mahler, \textit{Quantum Thermodynamics: Emergence of Thermodynamic Behavior Within Composite Quantum Systems (Lecture Notes in Physics)} (Springer, Berlin Heidelberg, 2009).
{\bibitem{Nori0}H.T. Quan, Y.D. Wang, Y.X. Liu, C.P. Sun, F. Nori,
Phys. Rev. Lett. \textbf{97}, 180402 (2006).}
{\bibitem{Nori} H. T. Quan, Y. X. Liu, C. P. Sun, and F. Nori, Phys. Rev. E \textbf{76}, 031105 (2007).}
{\bibitem{Nori1}K. Maruyama, F. Nori, V. Vedral, Rev. Mod. Phys. \textbf{81}, 1 (2009).}
{\bibitem{Brunner} N. Brunner, M. Huber, N. Linden, S. Popescu, R. Silva, and P. Skrzypczyk, Phys. Rev. E \textbf{89}, 032115 (2014).}
{\bibitem{Chotorlishvili} L. Chotorlishvili, Z. Toklikishvili, and J. Berakdar, J. Phys. A: Math. Theor. \textbf{44}, 165303 (2011).}
\bibitem{Uzdin} R. Uzdin, A. Levy, and R. Kosloff, Phys. Rev. X \textbf{5}, 031044 (2015).
\bibitem{Lostaglio} M. Lostaglio, K. Korzekwa, D. Jennings, and T. Rudolph, Phys. Rev. X \textbf{5}, 021001 (2015).
\bibitem{Goold} J. Goold, M. Huber, A. Riera, L. del Rio, and P. Skrzypczyk, J. Phys. A: Math. Theor. \textbf{49}, 143001 (2016).

\bibitem{Kosloff14} R. Kosloff and A. Levy, Annu. Rev. Phys. Chem. \textbf{65}, 365 (2014).
\bibitem{Gelbwaser15} D. Gelbwaser-Klimovsky, W. Niedenzu, and G. Kurizki, Adv. At. Mol. Phys. \textbf{64}, 329 (2015).
\bibitem{Alicki} R. Alicki, J. Phys. A: Math. Gen. \textbf{12}, L103 (1979).
\bibitem{Kosloff13} R. Kosloff, Entropy \textbf{15}, 2100 (2013).
\bibitem{Breuer} H. P. Breuer and F. Petruccione, \textit{The Theory of Open Quantum Systems} (Oxford University Press, New York, 2002).

\bibitem{Huelga} S. Huelga and M. Plenio, Contemp. Phys. \textbf{54}, 181 (2013).
\bibitem{Engel} G. S. Engel, T. R. Calhoun, E. L. Read, T-K. Ahn, T. Man\v{c}al, Y-C. Cheng, R. E. Blankenship, and G. R. Fleming, nature \textbf{446}, 782 (2007).

\bibitem{Gelbwaser} D. Gelbwaser-Klimovsky and A. Aspuru-Guzik, J. Phys. Chem. Lett. \textbf{6}, 3477 (2015).
\bibitem{Strasberg} P. Strasberg, G. Schaller, N. Lambert, and T. Brandes, New. J. Phys. \textbf{18}, 073007 (2016).
\bibitem{Newman} D. Newman, F. Mintert, and A. Nazir, arXiv:1609.04035.
\bibitem{Carrega} M. Carrega, P. Solinas, M. Sassetti, and U. Weiss, Phys. Rev. Lett. \textbf{116}, 240403 (2016).
\bibitem{Esposito15PRL} M. Esposito, M. A. Ochoa, and M. Galperin, Phys. Rev. Lett. \textbf{114}, 080602 (2015).
\bibitem{Esposito15PRB} M. Esposito, M. A. Ochoa, and M. Galperin, Phys. Rev. B \textbf{92}, 235440 (2015).
{\bibitem{Nitzan} A. Bruch, M. Thomas, S. V. Kusminskiy, F. von Oppen, and A. Nitzan, Phys. Rev. B \textbf{93}, 115318 (2016).}
\bibitem{Schmidt} R. Schmidt, M. F. Carusela, J. P. Pekola, S. Suomela, and J. Ankerhold, Phys. Rev. B \textbf{91}, 224303 (2015).

\bibitem{Tanimura88} Y. Tanimura and R. Kubo, J. Phys. Soc. Jpn. \textbf{58}, 101 (1988).
\bibitem{Ishizaki05} A. Ishizaki and Y. Tanimura, J. Phys. Soc. Jpn. \textbf{74}, 3131 (2005).
\bibitem{Tanimura06} Y. Tanimura, J. Phys. Soc. Jpn. \textbf{75}, 082001 (2006).
\bibitem{Tanimura14} Y. Tanimura, J. Chem. Phys. \textbf{141}, 044114 (2014).
\bibitem{Tanimura15} Y. Tanimura, J. Chem. Phys. \textbf{142}, 144110 (2015).
\bibitem{Kato15} A. Kato and Y. Tanimura, J. Chem. Phys, \textbf{143}, 064107(2015).
{\bibitem{Wu} J. Wu and J. Cao, J. Chem. Phys. \textbf{139}, 044102 (2013).}
%
{\bibitem{Tannor} E. Boukobza and D. J. Tannor, Phys. Rev. A \textbf{74}, 063823 (2006).}
{\bibitem{Segal} L. A. Wu and D. Segal, J. Phys. A: Math. Theor. \textbf{42}, 025302 (2009).}
{\bibitem{Castro} H. Hossein-Nejad, E. J. O'Reilly, and A. Olaya-Castro, New. J. Phys. \textbf{17}, 075014 (2015).}
{\bibitem{Schwinger} J. Schwinger, J. Chem. Phys. \textbf{2}, 407 (1961).}

\bibitem{KramerFMO}C. Kreisbeck and T. Kramer,  J. Phys. Chem. Lett. \textbf{3}, 2828 (2012).
\bibitem{Nori12}J. Ma, Z. Sun, X. Wang, and F. Nori, Phys. Rev. A \textbf{85}, 062323 (2012).
\bibitem{TanakaJPSJ09}M. Tanaka and Y.Tanimura, J. Phys. Soc. Jpn. \textbf{78}, 073802 (2009).
\bibitem{YanBO12}J. J. Ding, R. X. Xu, Y. J. Yan, J. Chem. Phys. \textbf{136}, 224103 (2012).
\bibitem{Liu} H. Liu, L. Zhu, S. Bai, and Q. Shi, J. Chem. Phys. \textbf{140}, 134106 (2014).
\bibitem{TanimruaJCP12}Y. Tanimura, J. Chem. Phys. \textbf{137}, 22A550 (2012).

\bibitem{Segal05} D. Segal and A. Nitzan, Phys. Rev. Lett. \textbf{94}, 034301 (2005).
\bibitem{Thoss} K. A. Velizhanin, H. Wang, and M. Thoss, Chem. Phys. Lett. \textbf{460}, 325 (2008).
\bibitem{Ruokola} T. Ruokola and T. Ojanen, Phys. Rev. B \textbf{83}, 045417 (2011).
\bibitem{Saito} K. Saito and T. Kato, Phys. Rev. Lett. \textbf{111}, 214301 (2013).
\bibitem{Wang} C. Wang, J. Ren, and J. Cao, Sci. Rep. \textbf{5}, 11787 (2015).

\bibitem{Scovil} H. E. D. Scovil and E. O. Schulz-DuBois, Phys. Rev. Lett. \textbf{2}, 262 (1959).
\bibitem{Geva} E. Geva and R. Kosloff, J. Chem. Phys. \textbf{104}, 7681 (1996).
\bibitem{Correa} L. A. Correa, J. P. Palao, D. Alonso, and G. Adesso, Sci. Rep. \textbf{4}, 3949 (2014).
\bibitem{Xu} D. Xu, C. Wang, Y. Zhao, and J. Cao, New. J. Phys. \textbf{18}, 023003 (2016).

\bibitem{YanPade10A} J. Hu, R. X. Xu, Y. J. Yan, J. Chem. Phys. \textbf{133}, 101106 (2010).
\bibitem{YanPade10B} B. L. Tian, J. J. Ding, R. X. Xu, Y. J. Yan, J. Chem. Phys. \textbf{133}, 114112 (2010).
\bibitem{Hu} J. Hu, M. Luo, F. Jiang, R-X. Xu, and Y. Yan, J. Chem. Phys. \textbf{134}, 244106 (2011).

\bibitem{Ye} L. Ye, D. Hou, R. Wang, D. Cao, X. Zheng, and Y. Yan, Phys. Rev. B \textbf{90}, 165116 (2014).
\bibitem{Cerrillo} J. Cerrillo, M. Buser, and T. Brandes, arXiv:1606.05074.

\bibitem{Agarwalla} B. K. Agarwalla, U. Harbola, W. Hua, Y. Zhang, and S. Mukamel, J. Chem. Phys. \textbf{142}, 212445 (2015).
\bibitem{Gao1} Y. Gao and M. Galperin, J. Chem. Phys. \textbf{144}, 174113 (2016).
\bibitem{Gao2} Y. Gao and M. Galperin, J. Chem. Phys. \textbf{144}, 244106 (2016).

\bibitem{Deffner} S. Deffner and C. Jarzynski, Phys. Rev. X \textbf{3}, 041003 (2013).
\end{thebibliography}

\end{document}